\documentclass[prd,preprint,tightenlines,floatfix,showpacs,preprintnumbers,nofootinbib,eqsecnum]{revtex4-1}
\pdfoutput=1

\usepackage{graphicx}
\usepackage{multirow}
\usepackage[english]{babel}
\usepackage{amsmath}
\usepackage{graphicx}
\usepackage{color}
\usepackage[dvipsnames]{xcolor}
\usepackage{amssymb}
\usepackage{hyperref}
\usepackage{dcolumn}
\usepackage{bm}
\usepackage{braket}
\usepackage[normalem]{ulem}
\usepackage[T1]{fontenc} 
\usepackage[utf8]{inputenc}

\def\beq{\begin{eqnarray}}
\def\eeq{\end{eqnarray}}

\def\SO{\mathop{\rm SO}}

\def\SU{\mathop{\rm SU}}
\def\U{\rm U}

\hypersetup{colorlinks=true, linkcolor=ForestGreen, urlcolor=blue, citecolor=teal}

\begin{document}

\title{Can $E_8$ unification at low energies be consistent with proton decay?}

\author{Alfredo Aranda}
\email{fefo@ucol.mx}
\affiliation{Facultad de Ciencias-CUICBAS, Universidad de Colima, C.P.28045, Colima, M\'exico 01000, M\'exico}
\affiliation{Dual CP Institute of High Energy Physics, C.P. 28045, Colima, M\'exico}

\author{Francisco~J.~de~Anda}
\email{fran@tepaits.mx}
\affiliation{Tepatitl{\'a}n's Institute for Theoretical Studies, C.P. 47600, Jalisco, M{\'e}xico}
\affiliation{Dual CP Institute of High Energy Physics, C.P. 28045, Colima, M\'exico}

\author{Ant{\'o}nio~P.~Morais}
\email{aapmorais@ua.pt}
\affiliation{Departamento de F\'isica, Universidade de Aveiro, Campus de Santiago, 
3810-183 Aveiro, Portugal}
\affiliation{Centre  for  Research  and  Development  in  Mathematics  and  Applications  (CIDMA), Campus de Santiago, 
3810-183 Aveiro, Portugal}

\author{Roman~Pasechnik}
\email{Roman.Pasechnik@hep.lu.se}
\affiliation{Department of Physics, Lund University, 221 00 Lund, Sweden\vspace{0.5cm}}

\begin{abstract}
\vspace{0.5cm}A model is presented that achieves unification of the full Standard Model (SM) field content into a single superfield. It has $E_8$ as a gauge group and simple SUSY in ten spacetime dimensions. The extra dimensions are orbifolded such that they reduce the gauge symmetry directly to the SM one. At low energies, only the SM field content remains with viable unified Yukawa couplings. Full unification can be achieved at energies as low as $10^6\ \rm{GeV}$ with controlled proton decay.
\end{abstract}

\maketitle

\section{Introduction}
\label{Sect:intro}

The idea that all the interactions present in the Standard Model (and perhaps gravity) are low energy manifestations of a single {\it unified} interaction at (typically very) high energies -- generically dubbed unification -- has driven a significant amount of theoretical explorations so far. The original proposal was made quickly after the consolidation of the Standard Model (SM) as the most compelling theory for particle physics at experimentally accessible energy scales~\cite{Georgi:1974sy,Fritzsch:1974nn}.

For the past four decades a plethora of models and scenarios have been concocted and studied that formally accomplish such a unification, i.e.~a successful (formal) embedding of the SM gauge group and matter content into a larger gauge symmetry, with varying degrees of phenomenological success. One can schematically (thus in an oversimplifying manner) think of three historical phases present in the approaches taken. A first phase is where group theoretical arguments gave the initial motivation and pathways to the creation of specific (proto-)models containing the basic ingredients and ideas. A second phase concerns the detailed phenomenological connection to observables leading in particular to two emblematic results: proton decay is present in most grand unified theories (GUTs) which in turn gives rise to important constraints on the unification scale, and in order to achieve gauge coupling unification, the presence of global supersymmetry (SUSY) was basically established as a requirement. Lastly, the third phase is associated to the development of string theory, where unification models took a strong reinforcement and a comprehensive study of the possibilities has been -- and continues to be -- carried out to this day \cite{Lerche:1986cx,Ibanez:1987sn,Parr:2020oar}. These three phases, albeit somewhat historical, have in fact co-existed in some ways. The presence of {\it extra dimensions} (EDs) in string theory did (re)motivate the exploration of scenarios involving them, both within the context of string theory and outside of it, for example. The {\it required} presence of SUSY in string theory and its role in gauge coupling unification lent impulse to the possibility of a bridge connecting string motivated models with the low-energy SM. Nevertheless, the broad theoretical success of string theory carried a substantial weight in the general direction of exploration during the recent past, leading to very important results and proposals that await a clear and observable connection to low-energy results.  

Given this situation, one may wonder why should another a letter/paper on unification be written. Besides the fact that the idea of unification is {\it appealing} to a significant sector of physicists, it is very important to consider the fact that, at this moment, there is still the possibility that models based on quantum field theory alone can accomplish the job, and second, that there can exist interesting venues that may have been overlooked for not being consistent with the framework of string theory. The work presented in this letter is motivated by such a situation. A model is presented, based on quantum field theory, that unifies the three interactions present in the SM and its matter content into a single representation of the $E_8$ group. The model is an ${\cal N}=1$ Super Yang Mills $E_8$ theory with a single gauge superfield in {\bf 248} representation living in a 10d spacetime with a very particular orbifold. An important observation is that the presence of ten dimensions is not related to string theory (the reason for this number of dimensions is explained below) and the orbifold accomplishes the necessary breaking of the gauge symmetry down to the SM gauge group as well as the required SUSY breaking, thus leading to a potentially predictive and testable model. 

Before describing the model in more detail it is convenient to outline the {\it typical} issues that GUTs must face and that in some cases rule them out. As mentioned above, once a connection to reality is undertaken, proton decay is a big obstacle: in fact, the simplest models have been ruled out by the bound on the proton's lifetime. Other more complex models are still allowed, and the proton's lifetime is translated into a lower bound for the unification scale which is typically required to be very high\footnote{I.e.~very far from direct experimental access.}, O($10^{17 - 18}$~GeV). Observed flavour structures present challenges as well. One would expect that a successful GUT could explain or at least naturally describe the hierarchies observed in the fermion mass spectrum as well as their mixing angles. A GUT that includes a flavour symmetry within its structure, for example, would be very attractive. Also alluded before, an actual (close) unification of gauge couplings is important and this can be difficult to achieve {\it naturally} without SUSY. This in turn introduces the necessity of assuming (or explaining) SUSY breaking and its effect in low energy phenomenology. These are some of the general issues associated to most GUTs. In the particular case of $E_8$, there is an important specific issue related to the fact that it only contains real representations and the SM is a chiral theory, leading to a construction based on ED setups with orbifold compactification. 

The setup presented here accomplishes a low-energy gauge couplings' unification consistent with proton decay bounds. Second and third generation quark masses emerge at tree level while first generation quarks and lepton masses are generated radiatively, thus inducing a strong hierarchy among them. Our theoretical considerations demonstrate that an appropriately chosen orbifold, including its associated Wilson lines, breaks the $E_8$ symmetry directly to the SM, leaving its particle content as the low-energy anomaly-free spectrum. In addition, we identify a few heavier fields possibly relevant to low energy phenomenology such as right-handed neutrinos, vector-like fermions and additional scalar doublets. The main aim of this Letter is to formulate such a novel $E_8$-based framework for Grand Unification model-building featuring the principal possibility for a strongly suppressed proton decay down to rather low unification scales. We leave a more comprehensive analysis of Wilson lines' breaking mechanism in the effective potential approach and the associated ED mechanism of SUSY breaking in this framework for a future dedicated and elaborate work.

\section{$E_8$ unification and breaking mechanism}
\label{Sect:orbifold}

The first question is why $E_8$ gauge group has been chosen for ultimate unification. The SM is defined by its gauge symmetry $G_{\rm{SM}} = \SU(3)_{\rm C}\times \SU(2)_{\rm L}\times \U(1)_{\rm Y}$ and its field content. The core idea of this model is to unify $G_{\rm{SM}} $ into a single symmetry group and to unify all fields into a single representation of it. The simplest way to achieve the gauge symmetry unification is with $\SU(5)$ \cite{Georgi:1974sy}. With	$\SO(10)$ one can unify all the SM fermions of a single family into one representation \cite{Fritzsch:1974nn}. In $E_6$, one could unify the SM Higgs sector and a full (but only one) family of fermions into a single representation \cite{King:2005my} by means of the simple ${\cal N}=1$ SUSY. The pattern of this enlarging of symmetry can be seen through the Dynkin diagrams to follow the exceptional chain \cite{Buchmuller:1985rc,Koca:1982zi},
\begin{equation}
 G_{\rm{SM}} \subset \SU(5) \subset \SO(10) \subset E_6 
 \subset E_7 \subset E_8 \,. \label{Eq:chain}
\end{equation}
For the largest group $E_8$, its adjoint representation $(\textbf{248})$ is the same as the fundamental one \cite{Slansky:1981yr}. This suggests that the full SM field content can be unified, provided that a maximal $\mathcal{N}=4$ SUSY is realised. Furthermore, it also provides an $\SU(3)$ flavour (or family) symmetry as a coset of $E_8$ to $E_6$ reduction.

Next, why should this occur in $10$d? A 10d spacetime is chosen because $\mathcal{N}=1$ SUSY in 10d decomposes into $\mathcal{N}=4$ SUSY in 4d, allowing full unification of the SM. Even though this also happens in 7d, 8d and 9d \cite{ArkaniHamed:2001tb}, only 10d allows to have the full SM with their required gauge representations. It would also be possible to build a similar model with more EDs but it would not be minimal\footnote{Note that these arguments have led to $E_8$ and a 10d spacetime from a completely different venue to that of string theory. This is certainly an interesting coincidence to be pointed out.}.

Six of the ten dimensions must form an orbifold suitable for a compactification that renders a realistic embedding of the SM. To do so, such an orbifold must accomplish the following: it should consistently break the $E_8$ symmetry down to $G_{\rm SM}$, distribute the $(\textbf{248})$ in such a way that the low-energy spectrum is that of the SM plus (perhaps) additional fields consistent with electroweak-scale phenomenology, and allow ${\cal N} = 1$ SYM in 10d while leading to a phenomenologically consistent low-energy theory in 4d. To do so, one can employ specific orbifold rotational boundary conditions as well as its continuous Wilson lines -- effective vacuum expectation values (VEVs) \cite{Ibanez:1986tp,Forste:2005rs}. It is the combination of these that leads to a potentially viable construction. While these techniques have been widely used for model building \cite{Aranda:2020noz, Aranda:2020fkj}, no known model has yet achieved a low-scale ultimate unification of matter and forces consistent with proton decay constraints. Exploring such a possibility is the main focus of this Letter.
\begin{figure}[h!]
	\centering
		\includegraphics[scale=0.4]{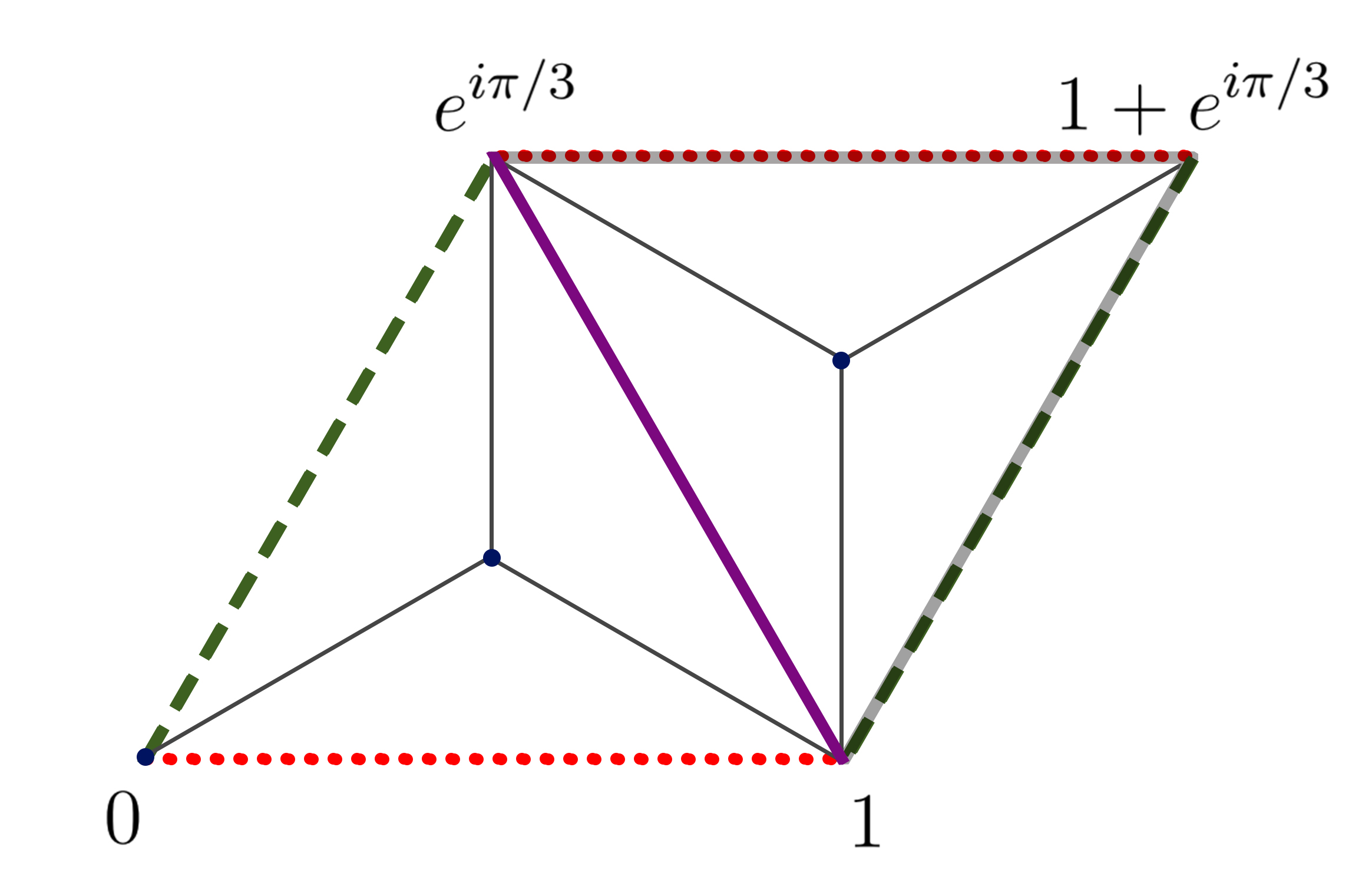}
		\includegraphics[scale=0.4]{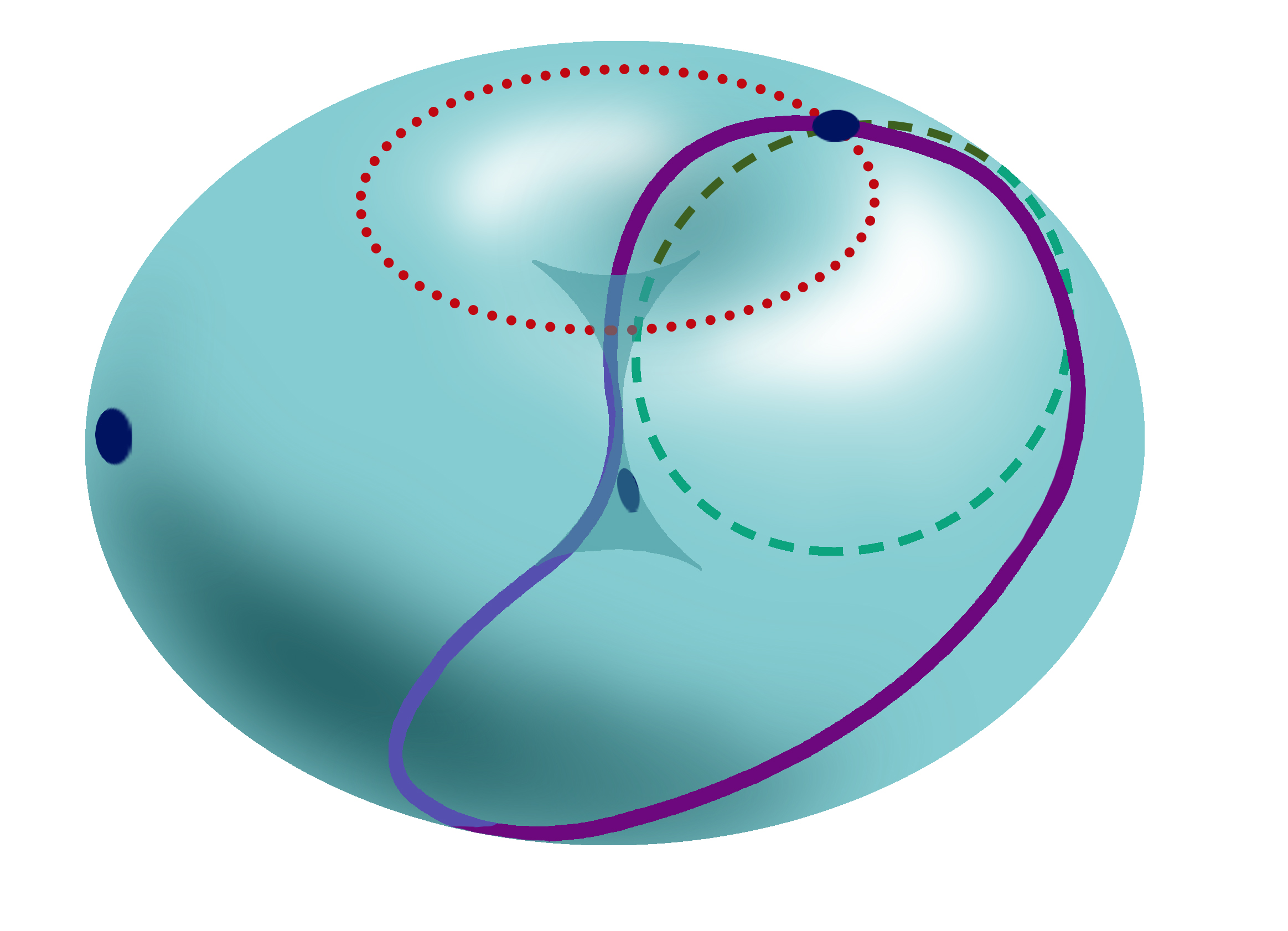}
		\caption{ The unfolded compactified ED space of a single coordinate $z_i$. To form the torus one identifies the green dotted lines together and the red dotted lines together. The coordinates are multiplied by the corresponding $R_i$.}
		\label{fig:toro}
\end{figure}
\begin{figure}[h!]
	\centering
 \begin{minipage}{0.6\textwidth}
\centering
		\includegraphics[scale=0.22]{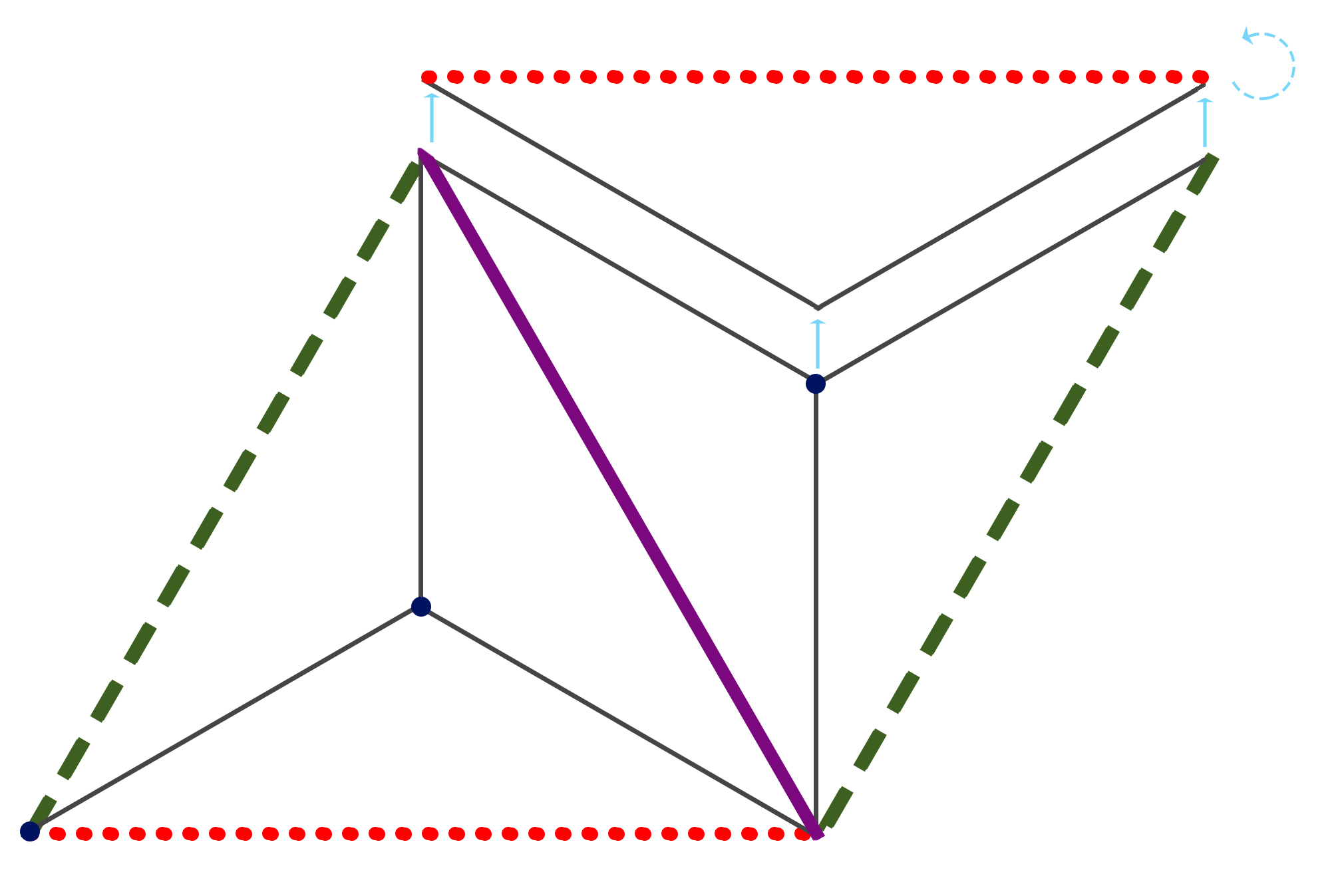}
		\includegraphics[scale=0.22]{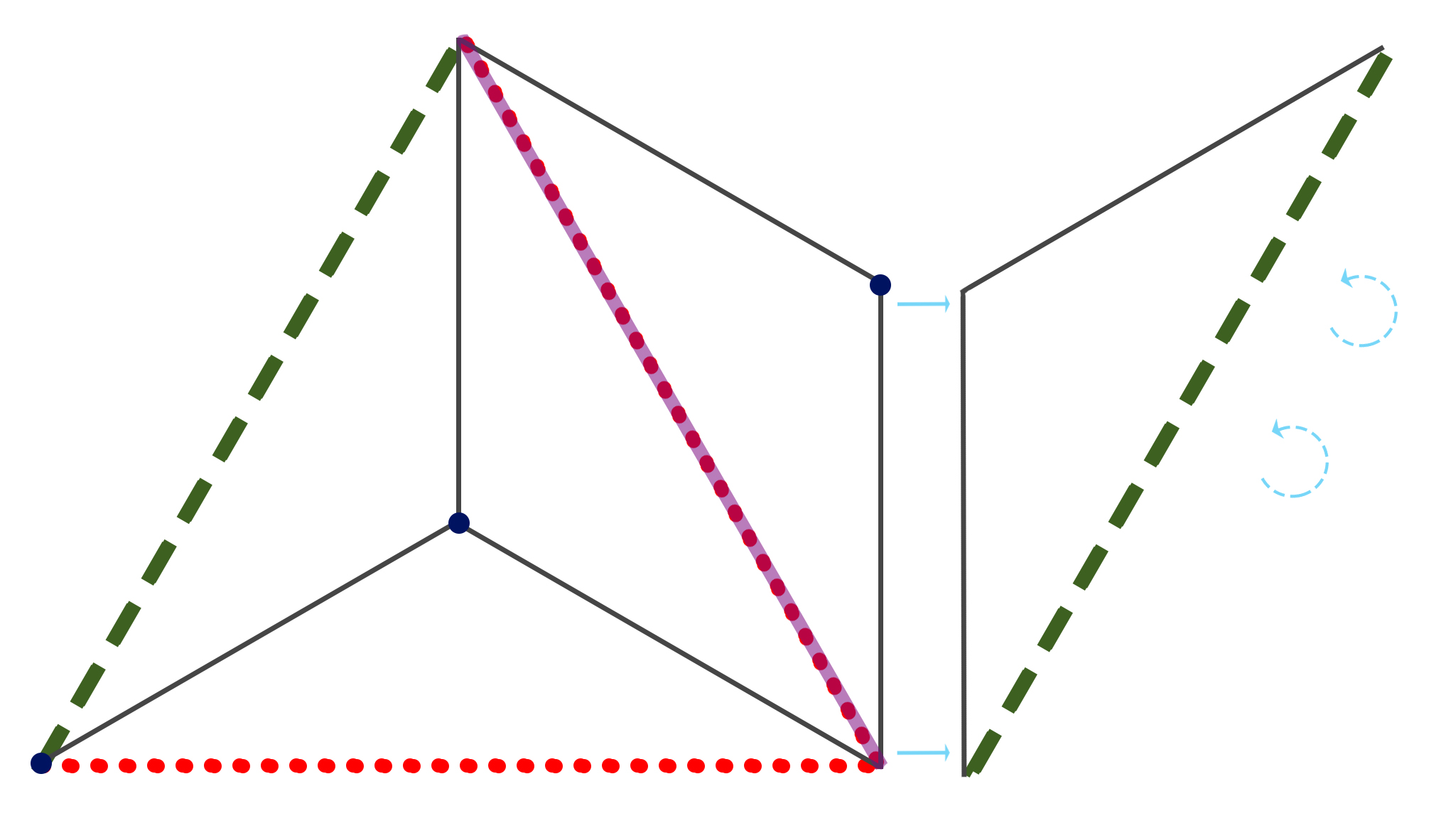}
		\includegraphics[scale=0.22]{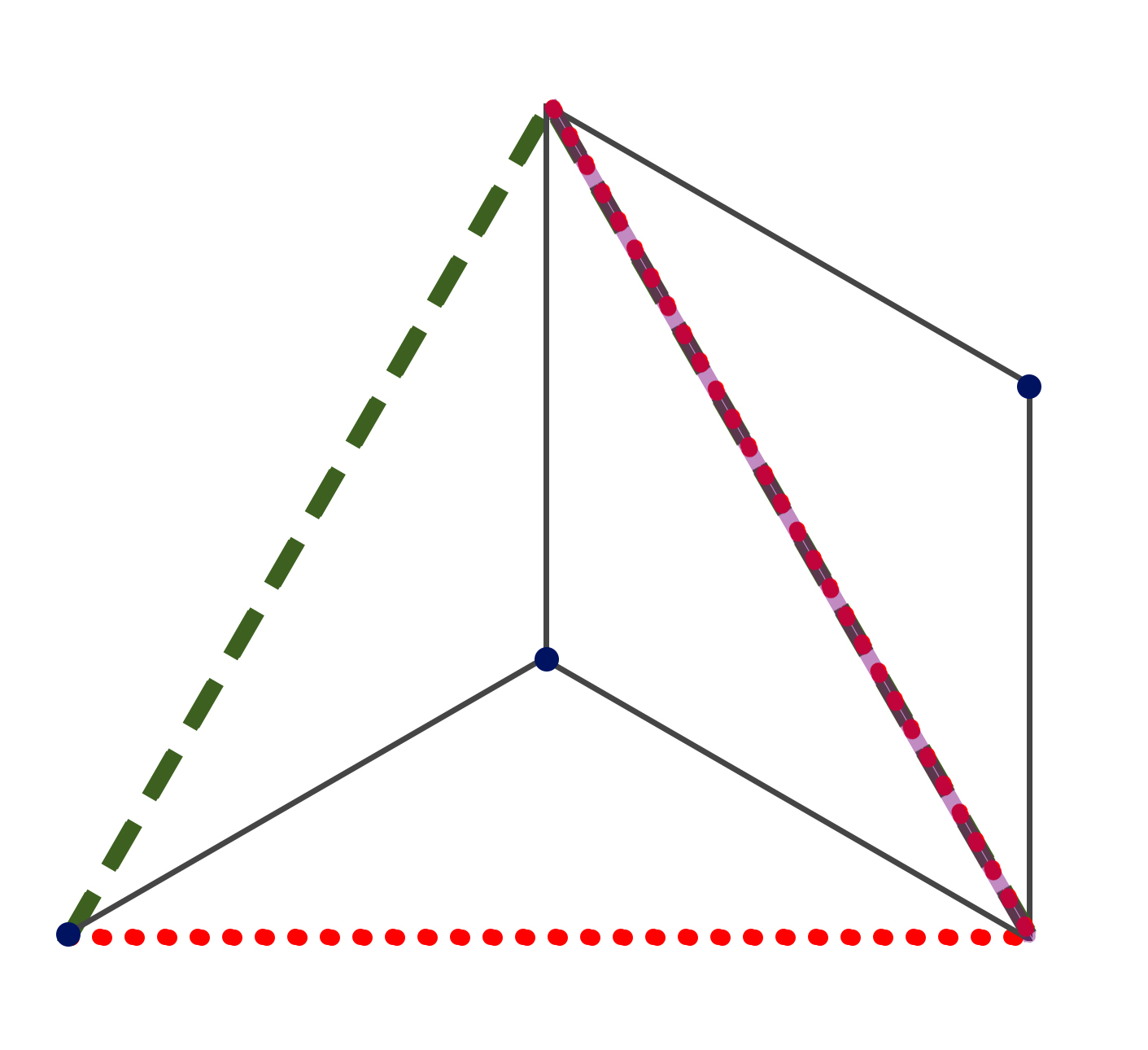}
		\includegraphics[scale=0.22]{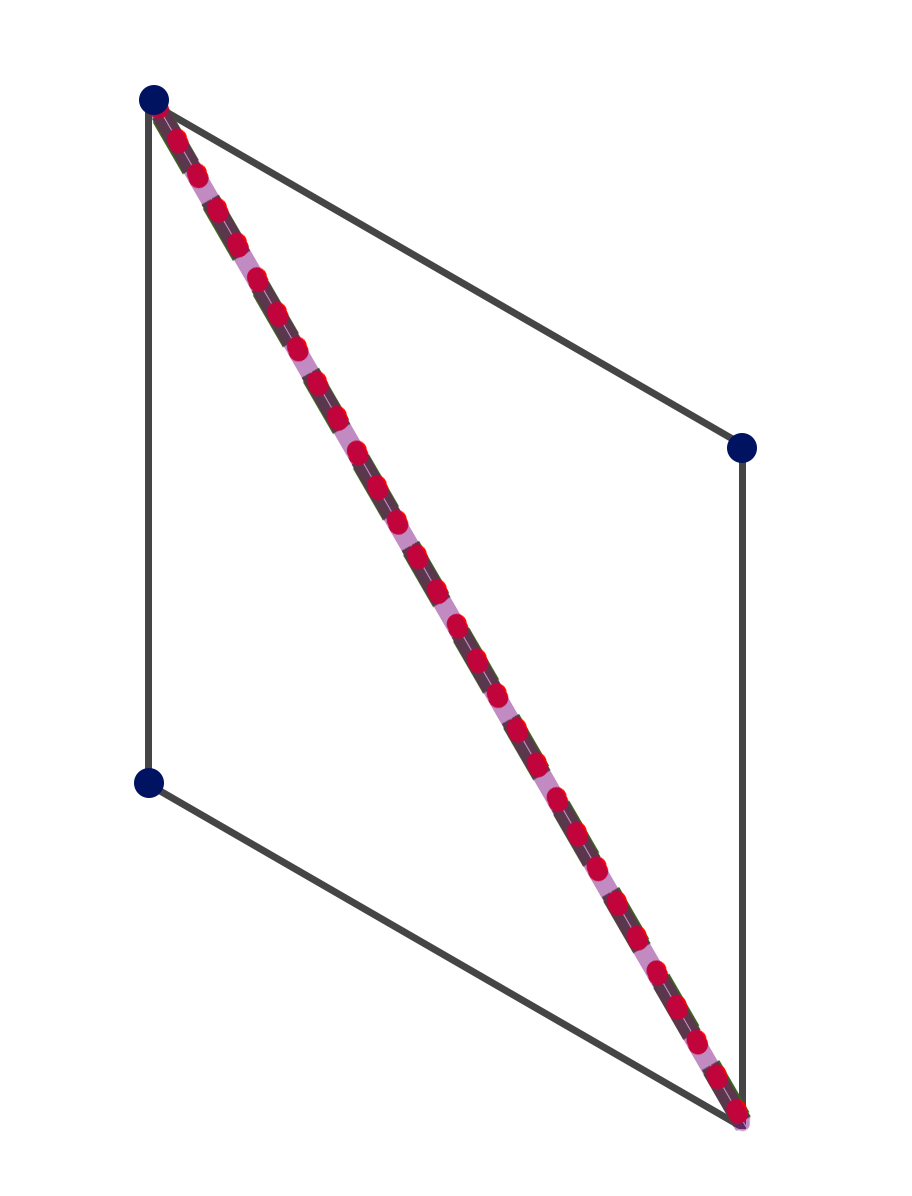}
         \includegraphics[scale=0.22]{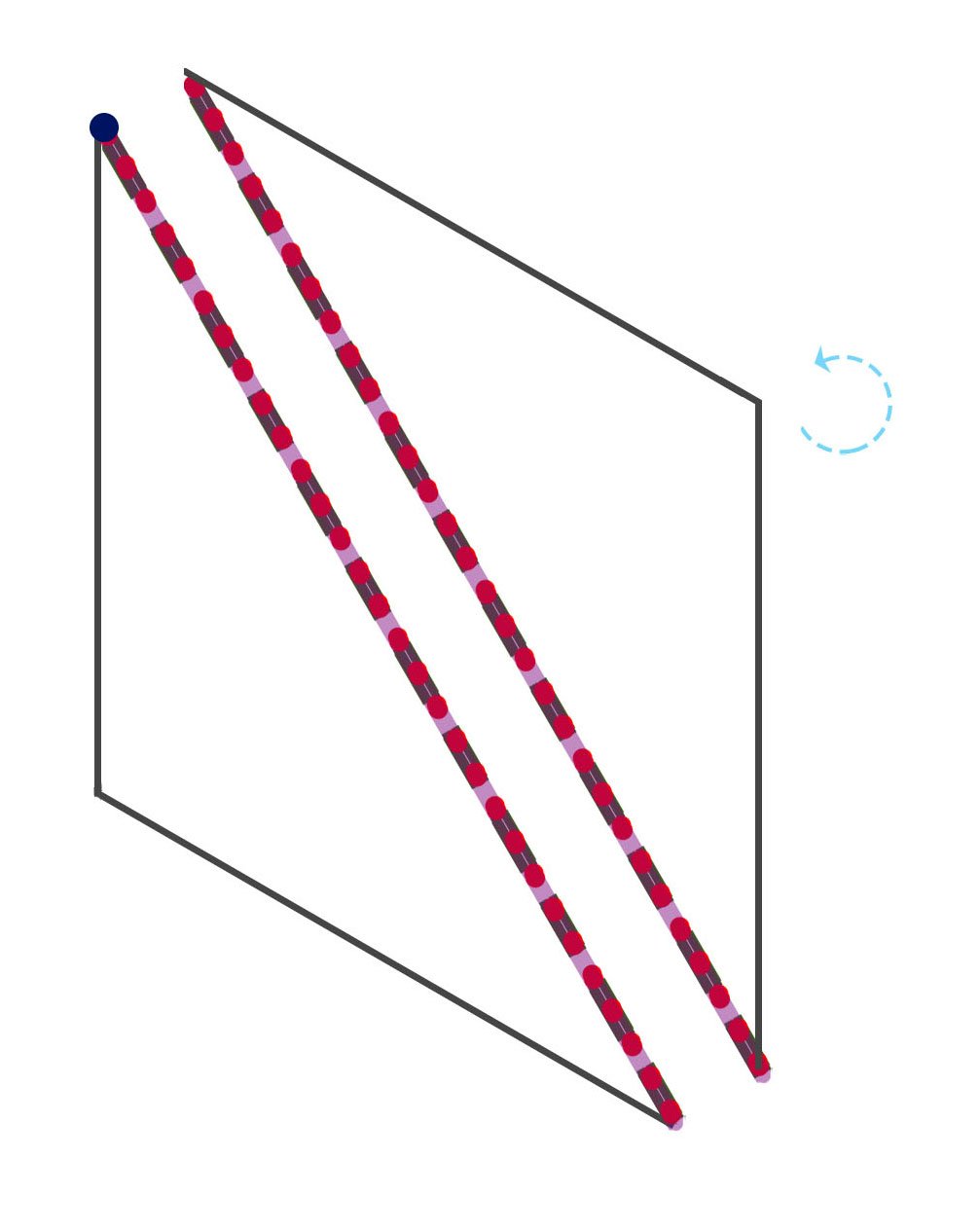}
         \includegraphics[scale=0.22]{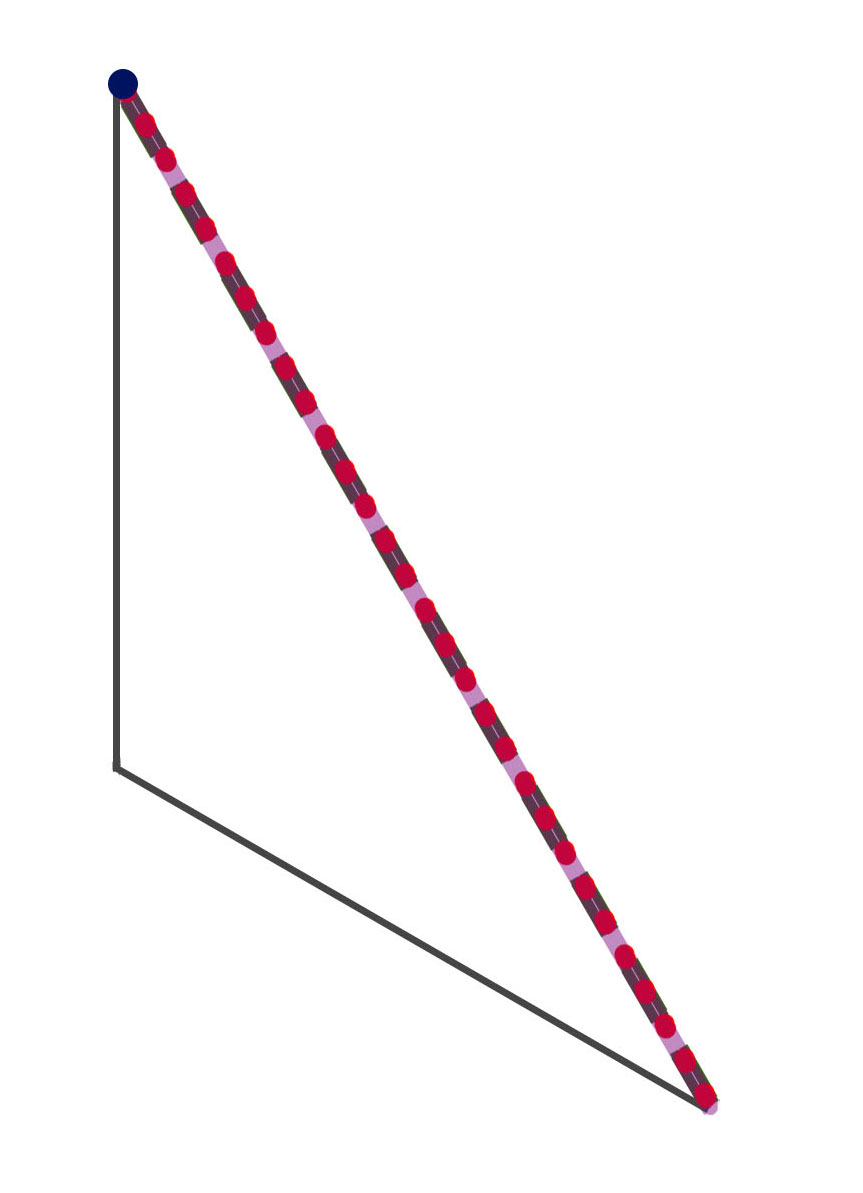}
\end{minipage}
 \begin{minipage}{0.39\textwidth}
\centering
		\includegraphics[scale=0.5]{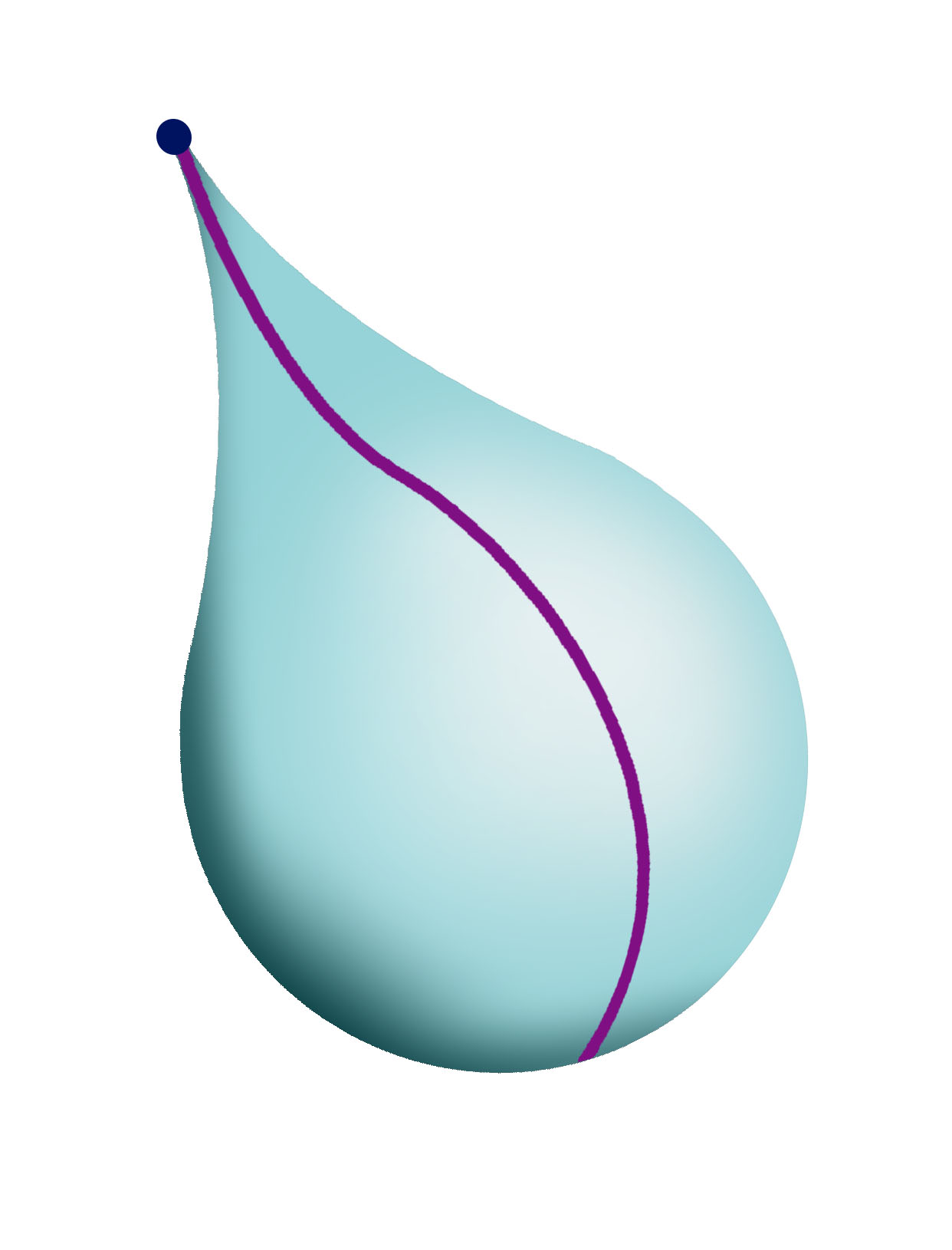}
\end{minipage}
		\caption{Folding the orbifold. The $\mathbb{Z}_6$ orbifolding corresponds to the following. First, one separates the whole space into two equilateral triangles. Then, each equilateral triangle is divided into three equal isosceles triangles, which are rotated and identified. Finally, the two remaining triangles are identified by reflecting one another. Note that the dotted lines are identified with the purple one.}
		\label{fig:folding-orbi}
\end{figure}
	
The considered model is an $\mathcal{N}=1$ SYM in 10d with the $E_8$ gauge group, constructed from a single 10d vector superfield $\mathcal{V}$ (built from a 10d real vector and a 10d Weyl/Majorana fermion) in the adjoint representation $(\textbf{248})$. The EDs (denoted by the three complex coordinates $z_i$) are orbifolded as $\mathbb{T}^6/(\mathbb{Z}_6\times \mathbb{Z}_2)$ with
\begin{equation} 
\begin{split}
\mathbb{Z}_6:\ (x,z_1,z_2,z_3)&\sim  (x,-\omega z_1,\omega z_2,-\omega z_3) \,, \qquad
\mathcal{V} \to e^{-2i\pi q^{X'}/6}\ \mathcal{V} \,, \\
\mathbb{Z}_2:\ (x,z_1,z_2,z_3)&\sim (x, -z_1, -z_2, z_3) \,, \qquad
\mathcal{V} \to e^{2i\pi (q^{Y}+q^F_8)/2}\ \mathcal{V} \,,
\label{eq:modrot}
\end{split}
\end{equation}
where $\omega=e^{2i\pi/3}$ and $q^*$ is the charge under the corresponding $\U(1)$ subgroup of $E_8$ \cite{Slansky:1981yr}. The $\mathbb{Z}_6$ breaks $E_8\to \SO(10)_{\rm GUT}\times \SU(3)_{\rm F}\times \U(1)_{\rm X'}$, while the $\mathbb{Z}_2$ breaks $E_8\to \SO(16)$ \cite{Aranda:2020noz,Aranda:2020fkj,Aranda:2020zms}\footnote{Note that $\SO(10)_{\rm GUT}\not\subset \SO(16)$.}. Together they break into the intersection $\SU(4)_{\rm PS}\times \SU(2)_{\rm L}\times \SU(2)_{\rm R}\times \SU(2)_{\rm F}\times \U(1)_{\rm X'}\times \U(1)_{\rm F8}$. The lattice of the $\mathbb{T}^6$ is
\begin{equation}
\begin{split}
z_i\sim z_i+2\pi R_i \,, \ \ \  z_{i}\sim z_{i} +2i\pi \omega^2 R_{i} \,,
\end{split}
\label{eq:modtra}
\end{equation}
where $R_i$ are the radii of three different tori. A continuous Wilson line is associated to $z_{1,2,3}$, whose gauge transformation is consistent with \eqref{eq:modrot}. The orbifolding process can be visualized (in a very schematic manner) in Figs.~\ref{fig:toro} and \ref{fig:folding-orbi}, where a 2d torus and a $\mathbb{Z}_6$ orbifold identifications are  shown. Choosing the three $R_i$ and the arbitrary dimensionless parameters associated to the continuous Wilson lines (where they are assumed to be aligned with the three right handed sneutrinos and two flavons) completely defines the model.

\section{Effective Pati-Salam theory}
\label{Sect:Pati-Salam}

The considered model is built on the orbifold defined by \eqref{eq:modrot}. Although this is the breaking associated to the rotational boundary conditions only (and the one associated to the Wilson line VEVs must also be considered), it is useful to use this intersection group as a basis for describing the $\textbf{248}$ decomposition into the corresponding modes. These are shown in Table~\ref{tab:pe3}, where the zero modes for each vector chiral multiplet are
{\small \begin{equation}\begin{split}
V_\mu &: \textcolor{magenta}{(\textbf{15},\textbf{1},\textbf{1},\textbf{1},0,0)}+\textcolor{magenta}{(\textbf{1},\textbf{3},\textbf{1},\textbf{1},0,0)}+\textcolor{magenta}{(\textbf{1},\textbf{1},\textbf{3},\textbf{1},0,0)}\\
&\ \ \ +\textcolor{magenta}{(\textbf{1},\textbf{1},\textbf{1},\textbf{3},0,0)}+\textcolor{magenta}{(\textbf{1},\textbf{1},\textbf{1},\textbf{1},0)}+\textcolor{magenta}{(\textbf{1},\textbf{1},\textbf{1},\textbf{1},0)}\\
\rightarrow & : \textcolor{magenta}{\bf G_{PS}}+\textcolor{magenta}{\bf W_L}+\textcolor{magenta}{\bf W_R}+\textcolor{magenta}{\bf W_F}+\textcolor{magenta}{\bf W_{F3}}+\textcolor{magenta}{\bf Z_{X'}}+\textcolor{magenta}{\bf Z_{F8}} \,, \\
\phi_1&:\textcolor{blue}{(\textbf{4},\textbf{2},\textbf{1},\textbf{2},1,1)} +\textcolor{blue}{(\bar{\textbf{4}},\textbf{1},\textbf{2},\textbf{1},1,-2)} \\
\rightarrow &: \textcolor{blue}{\bf F}+\textcolor{blue}{\bf f^c} \,, \\
\phi_2&: (\textbf{6},\textbf{1},\textbf{1},\textbf{2},-2,1)+\textcolor{OliveGreen}{(\textbf{1},\textbf{2},\textbf{2},\textbf{1},-2,-2)}+\textcolor{brown}{(\textbf{1},\textbf{1},\textbf{1},\textbf{2},4,1)} \\
\rightarrow &: \textbf{D}+\textcolor{OliveGreen}{\bf h}+\textcolor{brown}{\bf \Phi} \,,\\
\phi_3&:\textcolor{blue}{(\textbf{4},\textbf{2},\textbf{1},\textbf{1},1,-2)}+\textcolor{blue}{(\bar{\textbf{4}},\textbf{1},\textbf{2},\textbf{2},1,1)}\\
\rightarrow &: \textcolor{blue}{\bf f}+\textcolor{blue}{\bf F^c} \,.
\label{eq:zmt}
\end{split}\end{equation}}
\begin{table}[h]
	\centering
	\footnotesize
	\renewcommand{\arraystretch}{1.1}
	\begin{tabular}[t]{l|cccc}
		\hline
		 & $V$ & $\phi_1$ & $\phi_2$ & $\phi_3$\\ 
		\hline
	$\mathcal{V}_{\textcolor{magenta}{(\textbf{15},\textbf{1},\textbf{1},\textbf{1},0,0)}} $ & $1,1$ & $-\omega^2, -1$ & $\omega^2, -1$& $-\omega^2, 1$\\
		$\mathcal{V}_{\textcolor{magenta}{(\textbf{1},\textbf{3},\textbf{1},\textbf{1},0,0)}} $  & $1,1$ & $-\omega^2, -1$ & $\omega^2, -1$& $-\omega^2, 1$\\
		$\mathcal{V}_{\textcolor{magenta}{(\textbf{1},\textbf{1},\textbf{3},\textbf{1},0,0)}} $  & $1,1$ & $-\omega^2, -1$ & $\omega^2, -1$& $-\omega^2, 1$\\
		$\mathcal{V}_{\textcolor{magenta}{(\textbf{1},\textbf{1},\textbf{1},\textbf{1},0,0)}} $  & $1,1$ & $-\omega^2, -1$ & $\omega^2, -1$& $-\omega^2, 1$\\
		$\mathcal{V}_{\textcolor{magenta}{(\textbf{1},\textbf{1},\textbf{1},\textbf{3},0,0)}} $  & $1,1$ & $-\omega^2, -1$ & $\omega^2, -1$& $-\omega^2, 1$\\
		$\mathcal{V}_{\textcolor{magenta}{(\textbf{1},\textbf{1},\textbf{1},\textbf{1},0,0)}} $  & $1,1$ & $-\omega^2, -1$ & $\omega^2, -1$& $-\omega^2, 1$\\
		$\mathcal{V}_{\textcolor{brown}{(\textbf{1},\textbf{1},\textbf{1},\textbf{2},0,-3)}} $  & $1,1$ & $-\omega^2, -1$ & $\omega^2, -1$& $-\omega^2, 1$\\
		$\mathcal{V}_{\textcolor{brown}{(\textbf{1},\textbf{1},\textbf{1},\textbf{2},0,3)}} $ & $1,1$ & $-\omega^2, -1$ & $\omega^2, -1$& $-\omega^2, 1$\\
		$\mathcal{V}_{(\textbf{6},\textbf{2},\textbf{2},\textbf{1},0,0)} $ & $1,1$ & $-\omega^2, -1$ & $\omega^2, -1$& $-\omega^2, 1$\\
		$\mathcal{V}_{\textcolor{blue}{(\textbf{4},\textbf{2},\textbf{1},\textbf{1},-3,0)}} $  & $-1,-1$ & $\omega^2, 1$ & $-\omega^2, 1$& $\omega^2, -1$\\
		$\mathcal{V}_{\textcolor{blue}{(\bar{\textbf{4}},\textbf{1},\textbf{2},\textbf{1},-3,0)}} $ & $-1,1$ & $\omega^2, -1$ & $-\omega^2, -1$&  $\omega^2, 1$\\
		$\mathcal{V}_{\textcolor{red}{(\bar{\textbf{4}},\textbf{2},\textbf{1},\textbf{1},3,0)}} $  & $-1,-1$ & $\omega^2, 1$ & $-\omega^2, 1$& $\omega^2, -1$\\
		$\mathcal{V}_{\textcolor{red}{(\textbf{4},\textbf{1},\textbf{2},\textbf{1},3,0)}} $ & $-1,1$ & $\omega^2, -1$ & $-\omega^2, -1$&  $\omega^2, 1$\\
		$\mathcal{V}_{(\textbf{6},\textbf{1},\textbf{1},\textbf{2},-2,1)} $ & $\omega,-1$ & $-1,1$ & $1, 1$ & $-1, -1$\\
		$\mathcal{V}_{(\textbf{6},\textbf{1},\textbf{1},\textbf{1},-2,-2)} $ & $\omega,1$ & $-1,-1$ & $1, -1$ & $-1, 1$\\
		$\mathcal{V}_{(\textbf{6},\textbf{1},\textbf{1},\textbf{2},2,-1)} $  & $\omega^2,-1$ & $-\omega, 1$ &  $-1, 1$&  $-\omega, -1$\\
		$\mathcal{V}_{(\textbf{6},\textbf{1},\textbf{1},\textbf{1},2,2)} $  & $\omega^2,1$ & $-\omega, -1$ &  $-1, -1$&  $-\omega, 1$\\
		\hline
	\end{tabular} 
	\vspace{0.25cm}
	\begin{tabular}[t]{l|cccc}
		\hline
		 & $V$ & $\phi_1$ & $\phi_2$ & $\phi_3$\\ 
		\hline
$\mathcal{V}_{\textcolor{blue}{(\textbf{4},\textbf{2},\textbf{1},\textbf{2},1,1)}} $ & $-\omega,1$ & $1, -1$ &  $-1, -1$&  $1, 1$\\
	$\mathcal{V}_{\textcolor{blue}{(\textbf{4},\textbf{2},\textbf{1},\textbf{1},1,-2)}} $ & $-\omega,-1$ &  $1, 1$&  $-1, 1$&  $1, -1$\\
		$\mathcal{V}_{\textcolor{blue}{(\bar{\textbf{4}},\textbf{1},\textbf{2},\textbf{2},1,1)}} $ & $-\omega,-1$ &  $1, 1$&  $-1, 1$&  $1, -1$\\
		$\mathcal{V}_{\textcolor{blue}{(\bar{\textbf{4}},\textbf{1},\textbf{2},\textbf{1},1,-2)}} $ & $-\omega,1$ & $1, -1$ &  $-1, -1$&  $1, 1$\\
		$\mathcal{V}_{\textcolor{OliveGreen}{(\textbf{1},\textbf{2},\textbf{2},\textbf{2},-2,1)}} $  & $\omega,1$ &  $-1, -1$&  $1, -1$ & $-1, 1$\\
		$\mathcal{V}_{\textcolor{OliveGreen}{(\textbf{1},\textbf{2},\textbf{2},\textbf{1},-2,-2)}} $  & $\omega,-1$ & $-1, 1$ & $1, 1$&  $-1, -1$\\
		$\mathcal{V}_{\textcolor{brown}{(\textbf{1},\textbf{1},\textbf{1},\textbf{2},4,1)}} $ & $\omega,-1$ & $-1, 1$ & $1, 1$&  $-1, -1$\\
		$\mathcal{V}_{\textcolor{brown}{(\textbf{1},\textbf{1},\textbf{1},\textbf{1},4,-2)}} $ & $\omega,1$ &  $-1, -1$&  $1, -1$ & $-1, 1$\\
		$\mathcal{V}_{\textcolor{red}{(\bar{\textbf{4}},\textbf{2},\textbf{1},\textbf{2},-1,-1)}} $ & $-\omega^2,1$ & $\omega, -1$&  $-\omega, -1$&  $\omega, 1$ \\
		$\mathcal{V}_{\textcolor{red}{(\bar{\textbf{4}},\textbf{2},\textbf{1},\textbf{1},-1,2)}} $ & $-\omega^2,-1$ & $\omega, 1$&  $-\omega, 1$&  $\omega, -1$ \\
		$\mathcal{V}_{\textcolor{red}{(\textbf{4},\textbf{1},\textbf{2},\textbf{2},-1,-1) }} $  & $-\omega^2,-1$ & $\omega, 1$&  $-\omega, 1$&  $\omega, -1$ \\
		$\mathcal{V}_{\textcolor{red}{(\textbf{4},\textbf{1},\textbf{2},\textbf{1},-1,2)} } $  & $-\omega^2,1$ & $\omega, -1$&  $-\omega, -1$&  $\omega, 1$ \\
		$\mathcal{V}_{\textcolor{OliveGreen}{(\textbf{1},\textbf{2},\textbf{2},\textbf{2},2,-1)}} $ & $\omega,1$ &  $-1, -1$&  $1, -1$ & $-1, 1$\\
		$\mathcal{V}_{\textcolor{OliveGreen}{(\textbf{1},\textbf{2},\textbf{2},\textbf{1},2,2)}} $ & $\omega,-1$ & $-1, 1$ & $1, 1$&  $-1, -1$\\
		$\mathcal{V}_{\textcolor{brown}{(\textbf{1},\textbf{1},\textbf{1},\textbf{2},-4,-1)}} $  & $\omega,-1$ & $-1, 1$ & $1, 1$&  $-1, -1$\\
		$\mathcal{V}_{\textcolor{brown}{(\textbf{1},\textbf{1},\textbf{1},\textbf{1},-4,2)}} $  & $-\omega^2,1$ & $\omega, -1$&  $-\omega, -1$&  $\omega, 1$ \\
		\hline
	\end{tabular}
	\caption{Charges of each $\mathcal{N}=1$ superfield under $\SU(4)_{\rm PS}\times \SU(2)_{\rm L}\times \SU(2)_{\rm R}\times \SU(2)_{\rm F}\times \U(1)_{\rm X'}\times \U(1)_{\rm F8}$ including the orbifold ones. Only the fields with both orbifold charges equal to unity have zero modes. The representations are color coded as \textcolor{magenta}{adjoint fields}, \textcolor{ForestGreen}{Higgs and leptons}, \textcolor{blue}{quarks}, \textcolor{orange}{mirror Higgs and mirror fermions}, \textcolor{red}{mirror quarks}, and {\bf exotics}.} 
	\label{tab:pe3}
\end{table}
This is the anomaly-free field content that the purely ED space rotational boundary conditions leave massless. One still needs to introduce a mass splitting due to the Wilson line effective VEVs. The field content can be split into their components 
\begin{equation}
\begin{array}{lll}
\textcolor{blue}{\bf F}=(\textcolor{blue}{Q_{2,3}},\textcolor{blue}{L_{2,3}}), & \textcolor{blue}{\bf f}=(\textcolor{blue}{Q_{1}},\textcolor{blue}{L_{1}}) \,, & \\
\textcolor{blue}{\bf F^c}=\left(\begin{array}{cc}\textcolor{blue}{u^c_{2,3}} & \textcolor{violet}{\nu^c_{2,3}}\\ \textcolor{blue}{d^c_{2,3}} & \textcolor{blue}{e^c_{2,3}} \end{array}\right)\,, & \textcolor{blue}{\bf f^c}=\left(\begin{array}{cc}\textcolor{blue}{u^c_{1}} & \textcolor{violet}{\nu^c_{1}}\\ \textcolor{blue}{d^c_{1}} & \textcolor{blue}{e^c_{1}} \end{array}\right)\,, &\textcolor{OliveGreen}{\bf h}=\left(\begin{array}{c}\textcolor{OliveGreen}{h_u} \\ \textcolor{OliveGreen}{h_d} \end{array}\right) \,,\\
 \textbf{D}=(D_{2,3},D^c_{2,3}), & \textcolor{brown}{\bf \Phi}=\textcolor{brown}{\phi_{2,3}} \,,&
\end{array}
\end{equation}
where $\SU(3)_{\rm C}$ color indices are not shown. The assumed Wilson line aligned in \textcolor{violet}{$\nu^c_{1,2,3}$}, \textcolor{brown}{$\phi_{3}$} breaks 
\[
\SU(4)_{\rm PS}\times \SU(2)_{\rm L}\times \SU(2)_{\rm R}\times \SU(2)_{\rm F}\times \U(1)_{\rm X'}\times \U(1)_{\rm F8}\to \SU(3)_{\rm C}\times \SU(2)_{\rm L}\times \U(1)_{\rm Y} \,,
\]
leaving the following field content at low energies:
{\small
\begin{equation}
\begin{split}
   V_\mu :& \  \textcolor{magenta}{(\textbf{8},\textbf{1},0)}+\textcolor{magenta}{(\textbf{1},\textbf{3},0)}+\textcolor{magenta}{(\textbf{1},\textbf{1},0)},\\ 
   \phi_1 :& \ 2\times\textcolor{blue}{(\textbf{3},\textbf{2},1)}+2\times\textcolor{blue}{(\textbf{1},\textbf{2},-3)}+\textcolor{blue}{(\textbf{1},\textbf{1},6)} + \textcolor{violet}{(\textbf{1},\textbf{1},0)}+\textcolor{blue}{(\bar{\textbf{3}},\textbf{1},-4)}+\textcolor{blue}{(\bar{\textbf{3}},\textbf{1},2)}\\
   \phi_2: & \ \textcolor{OliveGreen}{(\textbf{1},\textbf{2},3)}+\textcolor{OliveGreen}{(\textbf{1},\textbf{2},-3)}+2\times \textcolor{brown}{(\textbf{1},\textbf{1},0)}+2\times(\bar{\textbf{3}},\textbf{1},2)+2\times (\textbf{3},\textbf{1},-2) \label{eq:zmff3} \\
   \phi_3:& \ \textcolor{blue}{(\textbf{3},\textbf{2},1)}+\textcolor{blue}{(\textbf{1},\textbf{2},-3)}+2\times\textcolor{blue}{(\textbf{1},\textbf{1},6)} +2\times \textcolor{violet}{(\textbf{1},\textbf{1},0)}+2\times\textcolor{blue}{(\bar{\textbf{3}},\textbf{1},-4)}+2\times\textcolor{blue}{(\bar{\textbf{3}},\textbf{1},2)} \,.
\end{split}
\end{equation}}
denoted as 
\begin{equation}\begin{split}
V_\mu &:  \textcolor{magenta}{G_\mu}+\textcolor{magenta}{W_\mu}+\textcolor{magenta}{B_\mu} \,,\\
\phi_1&:\textcolor{blue}{Q_{2,3}}+\textcolor{blue}{L_{2,3}}+\textcolor{blue}{e^c_1}+ \textcolor{violet}{\nu^c_1}+\textcolor{blue}{u^c_1}+\textcolor{blue}{d^c_1}\\
\phi_2&:\textcolor{OliveGreen}{h_{u}}+\textcolor{OliveGreen}{h_{d}}+\textcolor{brown}{\varphi_{2,3}}+D^c_{2,3}+D_{2,3} \,,\\
\phi_3&: \textcolor{blue}{Q_1}+\textcolor{blue}{L_1}+\textcolor{blue}{e^c_{2,3}}+ \textcolor{violet}{\nu^c_{2,3}}+\textcolor{blue}{u^c_{2,3}}+\textcolor{blue}{d^c_{2,3}} \,,
\end{split}
\label{eq:0fields}
\end{equation}
respectively. These are the fields that remain in the theory below the compactification scale -- the massless SM fields plus three massive right-handed neutrinos as well as two Higgs doublets, two flavons, and two vector-like triplet pairs. The heavy fields, such as the vector-like weak-triplet and singlet fermions, acquire their masses via the effective VEVs at high scales.

In this model, one must build effective Lagrangian terms from the fields in \eqref{eq:zmt}. They must respect the symmetry $\SU(4)_{\rm PS}\times \SU(2)_{\rm L}\times \SU(2)_{\rm R}\times \SU(2)_{\rm F}\times \U(1)_{\rm X'}\times \U(1)_{\rm F8}$ and only be broken through the effective VEVs in $\braket{\textcolor{brown}{\varphi_3}},\braket{\textcolor{violet}{\nu^c_{1,2,3}}}$. Furthermore, higher-order operators must be mediated by KK modes. At leading order, the gauge-field strength superpotential and the K\"ahler potential are the standard ones, and there is a unique superpotential term \cite{Marcus:1983wb},
\begin{equation}
\begin{split}
    \mathcal{W}\sim& g \phi_1\phi_2\phi_3\sim \textcolor{blue}{F} \textcolor{OliveGreen}{h} \textcolor{blue}{F^c}+D\big[\textcolor{blue}{Ff} + \textcolor{blue}{F^cf^c}\big]\\
    &=\textcolor{OliveGreen}{h_u}(\textcolor{blue}{L_{2,3}}\textcolor{violet}{\nu^c_{2,3}}+\textcolor{blue}{Q_{2,3}}\textcolor{blue}{u^c_{2,3}})+ \textcolor{OliveGreen}{h_d}(\textcolor{blue}{L_{2,3}}\textcolor{blue}{e^c_{2,3}}+\textcolor{blue}{Q_{2,3}}\textcolor{blue}{d^c_{2,3}})
    \\
    &\quad + D_{2,3}(\textcolor{blue}{Q_{2,3}}\textcolor{blue}{Q_1}+\textcolor{blue}{d^c_{2,3}}\textcolor{violet}{\nu^c_1}+\textcolor{blue}{d^c_{1}}\textcolor{violet}{\nu^c_{2,3}}+\textcolor{blue}{u^c_{2,3}}\textcolor{blue}{e^c_1}+\textcolor{blue}{u^c_{1}}\textcolor{blue}{e^c_{2,3}})
    \\
    &\quad
    +D^c_{2,3}(\textcolor{blue}{u^c_{2,3}}\textcolor{blue}{d^c_1}+\textcolor{blue}{d^c_{2,3}}\textcolor{blue}{u^c_1}+\textcolor{blue}{Q_{2,3}}\textcolor{blue}{L_{1}}+\textcolor{blue}{Q_{1}}\textcolor{blue}{L_{2,3}})
    \\ &\quad+D_{2,3}\textcolor{blue}{d^c_{2,3}}\braket{\textcolor{violet}{\nu^c_1}}+D_{2,3}\textcolor{blue}{d^c_{1}}\braket{\textcolor{violet}{\nu^c_{2,3}}}+\textcolor{OliveGreen}{h_u}\textcolor{blue}{L_{2,3}}\braket{\textcolor{violet}{\nu^c_{2,3}}} \,,
    \label{eq:pssup}
\end{split}
\end{equation}
where every term is proportional to the gauge coupling up to Clebsch-Gordan coefficients (CGC). The second line contains the second and third family Yukawa terms. The second and third lines contain quark couplings with the vector-like triplets. The last line includes \textcolor{blue}{$d^c$}-$\bar{D}$ mixing which gives masses to the extra triplets (and the light \textcolor{blue}{$d^c$} lies in a linear combination of them) and an R-parity violating \textcolor{OliveGreen}{$h$}-\textcolor{blue}{$L$} mixing which can be rotated into a $\mu$ term, while the R-parity violation can be bound within the experimental constraints \cite{Aranda:2021eyn,Dreiner:1997uz}. The right-handed neutrinos and flavon obtain a Majorana mass from the K\"ahler potential and its related effective VEVs. As within our conjecture, non-zero Wilson line VEVs would break SUSY by means of the $\mathcal{D}$-term as
\begin{eqnarray}
\braket{\mathcal{D}_A} = \sum_{ij} \braket{\textcolor{violet}{\nu^c_i}}^\dagger t_	A\braket{\textcolor{violet}{\nu^c_j}} + 
\braket{\textcolor{brown}{\varphi_3}}^\dagger t_A \braket{\textcolor{brown}{\varphi_3}}
+ \braket{\textcolor{brown}{\varphi_3}}^\dagger t_A\braket{\textcolor{violet}{\nu^c_j}} + 
\braket{\textcolor{violet}{\nu^c_i}}^\dagger t_A\braket{\textcolor{brown}{\varphi_3}}\neq 0 \,,
\end{eqnarray} 
where the fermion in \textcolor{brown}{$\varphi_2$} becomes a massless goldstino. The details of the SUSY breaking mechanism in the proposed framework are subject for further in-depth studies.

\section{Gauge couplings unification and proton decay}
\label{Sect:unification&proton}

The unification of gauge interactions can be demonstrated at one-loop level by means of the renormalization group (RG) evolution of the three SM gauge couplings $g_\mathrm{A},~\mathrm{A} = \U(1)_{\rm Y},~\SU(2)_{\rm L},~\SU(3)_{\rm C}$, from the electroweak up to the unification/compactification scale. To do so, the effect of all Kaluza-Klein (KK) modes must be incorporated. The three distinct radii of the orbifold provide different intermediate scales that must be incorporated into the RG running according to \cite{Dienes:1998vg,Aranda:2021eyn}. This results in a modification of the usual logarithmic evolution according to
\begin{equation}
    \beta^{(1)}_{g_\mathrm{A}} \to \beta^{(1)}_{g_\mathrm{A}} + \left[S(\mu,\delta) -1\right] \tilde{\beta}^{(1)}_{g_\mathrm{A}} \,.
     \label{eq:beta}
 \end{equation}
Here, the contribution of the KK modes is encoded in $\tilde{\beta}^{(1)}_{g_\mathrm{A}}$, and
\begin{equation}
     S(\mu,\delta) = \frac{2 \pi^{\delta/2}}{\delta ~ \Gamma(\delta/2)} \left(\frac{\mu}{\mu_\mathrm{KK}}\right)^\delta \qquad \text{for} \qquad \mu \geq \mu_\mathrm{KK}\,,
     \label{eq:power}
\end{equation}
with $\delta$ representing the number of EDs, $\mu_\mathrm{KK}$ being the scale at which the KK modes enter the particle spectrum, and $\Gamma(x)$ -- the Euler gamma function. 

\begin{figure}[h!]
	\centering
		\includegraphics[scale=1.0]{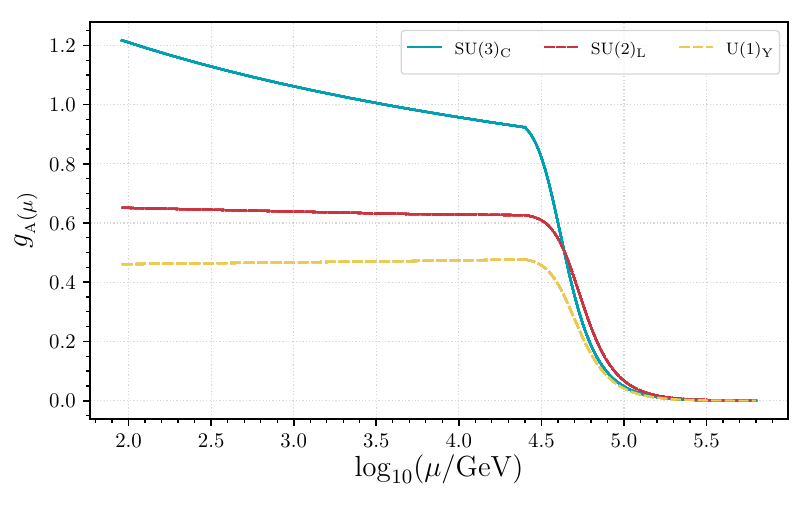}
		\caption{Generic pattern of the RG running of the three SM gauge couplings into their unification at a scale of $\Lambda_0 = 10^{5.8}$ GeV. See~\cite{Aranda:2021eyn} for further details.}
		\label{fig:gunification}
\end{figure}

One then defines the RG running taking into account the two distinct regions. In the first region, the contribution of the KK modes living in the $\delta=6$ EDs yields a power-law running as
\begin{equation}
    \tilde{\alpha}_\mathrm{A}^{-1}\left(\mu{}{}\right) = \alpha_{\mathrm{A}(1)}^{-1} - \tilde{b}_\mathrm{A} \frac{\pi^2}{72} \left[ \left( \frac{\mu}{\mu_\mathrm{KK}} \right)^6 - 1 \right]\,.
  \label{eq:RGE-pow}
\end{equation}
In the second region, the effect of the KK modes wears out resulting in the usual logarithmic running as
\begin{equation}
    \alpha_\mathrm{A}^{-1}\left(\mu{}{}\right) = \alpha_{\mathrm{A}(0)}^{-1} - \frac{b_\mathrm{A}}{2 \pi} \log\frac{\mu}{\mu_0}\,,
  \label{eq:RGE-log}
\end{equation}
with $\alpha_\mathrm{A}^{-1} \equiv (4 \pi) / g_\mathrm{A}^2$ representing the inverse structure constants\footnote{For simplicity of notation, one uses the inverse structure constants in Eqs.~\eqref{eq:RGE-pow} and \eqref{eq:RGE-log} while in Fig.~\ref{fig:gunification}, for a clearer visualization, one represents the evolution of the gauge couplings as such.}.

In the above equations the matching condition $\alpha_{\mathrm{A}(1)}^{-1} = \alpha_\mathrm{A}^{-1} (\mu_\mathrm{KK})$ is implied while $\alpha_{\mathrm{A}(0)}^{-1}$ denotes the measured values of the inverse structure constants at the electroweak scale $\mu_0 = M_Z$. According to the authors' previous discussion in \cite{Aranda:2021eyn}, the scale below which the KK modes wear out is defined in terms of the radii of the extra dimensions as $\mu_\mathrm{KK} = \left(R_1 R_2 R_3 \right)^{-1/3}$, where in this letter one chooses $R_1 < R_2 = R_3$, with $\Lambda_0 \equiv 1/R_1$ being the unification/compactification scale. Last but not least, $b_\mathrm{A}$ and $\tilde{b}_\mathrm{A}$ denote the coefficients of the beta-functions for the logarithmic and power-law running regions, respectively.

A specific example is shown in Fig.~\ref{fig:gunification} where the smallest unification scale was numerically found to be $\Lambda_0 = 10^{5.8}~\mathrm{GeV}$ in consistency with proton decay constraints (see a follow-up discussion below) and $\mu_\mathrm{KK} = 10^{4.4}~\mathrm{GeV}$. In the $\left[\mu_\mathrm{KK},\Lambda_0\right]$ region, taking into account the KK modes that correspond to the degrees of freedom specified in Eq.~\eqref{eq:zmt} (or in Tab.~\ref{tab:pe3}) and projected onto the SM gauge group, one obtains
\begin{equation}
    \tilde{b}_{\U(1)_{\rm Y}} = -13\,, \qquad \tilde{b}_{\SU(2)_{\rm L}} = -5\,, \qquad \tilde{b}_{\SU(3)_{\rm C}} = -10\,,
\end{equation}
where the presence of charged vectors results in the asymptotic safety of $\mathrm{U}(1)_\mathrm{Y}$. 

For the region where the RG running is logarithmic two sub-regions must be considered. The first one is between $\mu_\mathrm{KK}$ and $\Lambda_1 = 10^{3.7}~\mathrm{GeV}$, where the fields in Eqs.~\eqref{eq:zmff3} and \eqref{eq:0fields} are included, resulting in
\begin{equation}
    b_{\U(1)_{\rm Y}} = 149/30\,, \qquad b_{\SU(2)_{\rm L}} = -3/2\,, \qquad b_{\SU(3)_{\rm C}} = -7\,.
\end{equation}
In the second sub-region, i.e.~below $\Lambda_1$ and until the electroweak scale, only the SM fields are present yielding the usual SM beta-function coefficients,
\begin{equation}
    b_{\U(1)_{\rm Y}} = 41/10\,, \qquad b_{\SU(2)_{\rm L}} = -19/6\,, \qquad b_{\SU(3)_{\rm C}} = -7\,.
\end{equation}

The effect of the KK modes in Fig.~\ref{fig:gunification} is evident, leading to an asymptotic unification at a scale of O($10^6$~GeV) and a gauge coupling strength at a value of O($10^{-6}$). This scale is obtained solely by satisfying a safe proton lifetime and ED constraints as discussed below (see also~\citep{Aranda:2021eyn}). At this scale, the considered model becomes sensitive to constraints from Pati-Salam leptoquark masses to be $M_{\rm LQ}>1.2\ {\rm PeV}$ \cite{ParticleDataGroup:2022pth}. Notably, we reach an overall consistency with the PeV-scale SUSY breaking \cite{Wells:2004di}. Such a low unification scale is both interesting, since it has a better chance to be probed by measurements eventually than in more conventional GUTs, and dangerous. Indeed, the proton's lifetime provides a strong constraint on the unification scale of most GUTs. 

In order to see how the proposed setup leads to a strongly suppressed proton decay, first notice that the vector-like triplets $D,D^c$ cannot mediate it through the superpotential in Eq.~(\ref{eq:pssup}). Indeed, the latter only involves two second-family quarks $\sim$\textcolor{blue}{$Q_2 Q_1$}\textcolor{violet}{$\nu^c_j$}$D_2/\braket{\textcolor{violet}{\tilde{\nu}^c_k}}$, which are not related to proton decay. Furthermore, as the Pati-Salam gauge-vector fields do not mediate proton decay, there are no zero-mode proton decay mediators present in our spectrum. There are, however, $E_8$ fields that do, but they are all (non-zero) KK modes.

\begin{figure}[h!]
	\centering
		\includegraphics[scale=0.18]{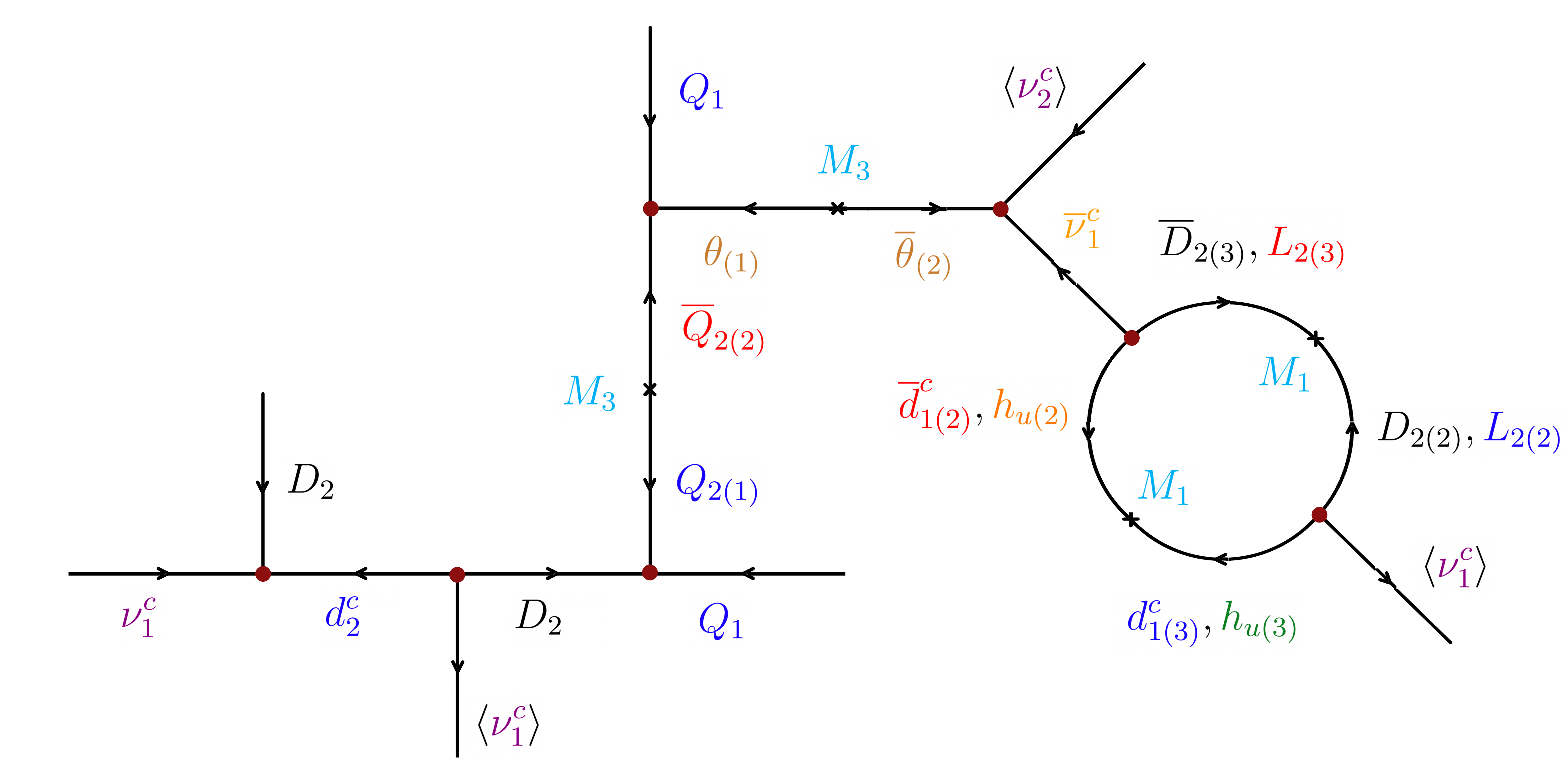}
		\caption{Dominant channel for proton decay. Each vertex (marked by a red dot) is $\sim g_8$. The internal lines with subindex $(j)$ denote KK modes coming from the chiral superfield $\phi_j$ and with mass $M_k\sim 1/R_k$. The barred fields lie in the mirror representation of the corresponding field. The flavon mediator is $\textcolor{brown}{\theta}\sim \textcolor{brown}{(\textbf{1},\textbf{1},\textbf{1},\textbf{2},0,3)}$.}
		\label{fig:pdd}
\end{figure}

In this model, proton decay is then generated at one-loop level, and the main channel involves the previously mentioned vector-like triplets $D,D^c$, and which requires the loop in order to change one family. The relevant term is shown in Fig.~\ref{fig:pdd}, where each vertex introduces a factor proportional to the universal $g_8$ gauge coupling of the high-scale $E_8$ theory. Hence, the corresponding amplitude is of order $\sim g_8^5$ provided that the $D$-\textcolor{blue}{$d^c$} mass enters in the denominator and takes away two factors of $g_8$. As will be illustrated below, the $g_8$ coupling is very small at the compactification scale, where such an operator is generated, naturally leading to a very strong suppression of the corresponding proton decay rate.

Likewise, the mirror KK-mode fermions may also mediate proton decay, with a diagram very similar to the previous one, and such mediators are bound to have mass at or beyond the compactification scale. Then, the corresponding decay rate estimate would be the same as in the previous case, but with $M_D\to M_X$, which leads to even stronger suppression. Finally, the KK gauge vectors inside $E_8$ that correspond to the non-SM $\SU(5)$ generators could also mediate proton decay. However, such contributions are always more suppressed than the one in Fig.~\ref{fig:pdd} due to a higher power of $g_8$ involved.

The dominant proton decay channel corresponding to the diagram in Fig.~\ref{fig:pdd} goes mainly into kaons, with the rate estimated as~\cite{Antusch:2014poa}
\begin{equation}
    \Gamma_{p\to K+L}\sim g_8^{10}l^4\braket{\textcolor{violet}{\tilde{\nu}^{c\dagger }_1\tilde{\nu}^c_2}}^2\frac{m_p^5}{M_{\rm KK}^{12}} \,,
    \label{eq:prodec}
\end{equation}
in terms of the loop suppression factor $l$ and the lightest KK mass at the GUT/compactification scale, $M_{\rm KK}$, while the flavor preserving channels are more suppressed \cite{Murayama:1994tc}. Such a decay rate is less constrained than the one into pions, $\tau_{p\to K+L}=6.6\times 10^{-3}\ \tau_{p\to \pi+L}$~\cite{Zyla:2020zbs}, with the most stringent bound on the proton lifetime $\tau_{p\to \pi+L}<1.7\times 10^{34}\ {\rm yrs}$~\cite{Takhistov:2016eqm}. These constraints set the bound on the GUT/compactification scale from Eq.~\eqref{eq:prodec} as
\begin{equation}
    \Lambda_0 \equiv M_{\rm KK}>g_8^{5/2}\ l\ \ 7.7\times 10^{15}\ {\rm GeV} \,,
    \label{eq:conspd}
\end{equation}
where the small $l,g_8\ll 1$ may bring it down by many orders of magnitude, thus, enabling low-scale Grand Unification in our context. 

\begin{figure}[h!]
	\centering
		\includegraphics[scale=0.6]{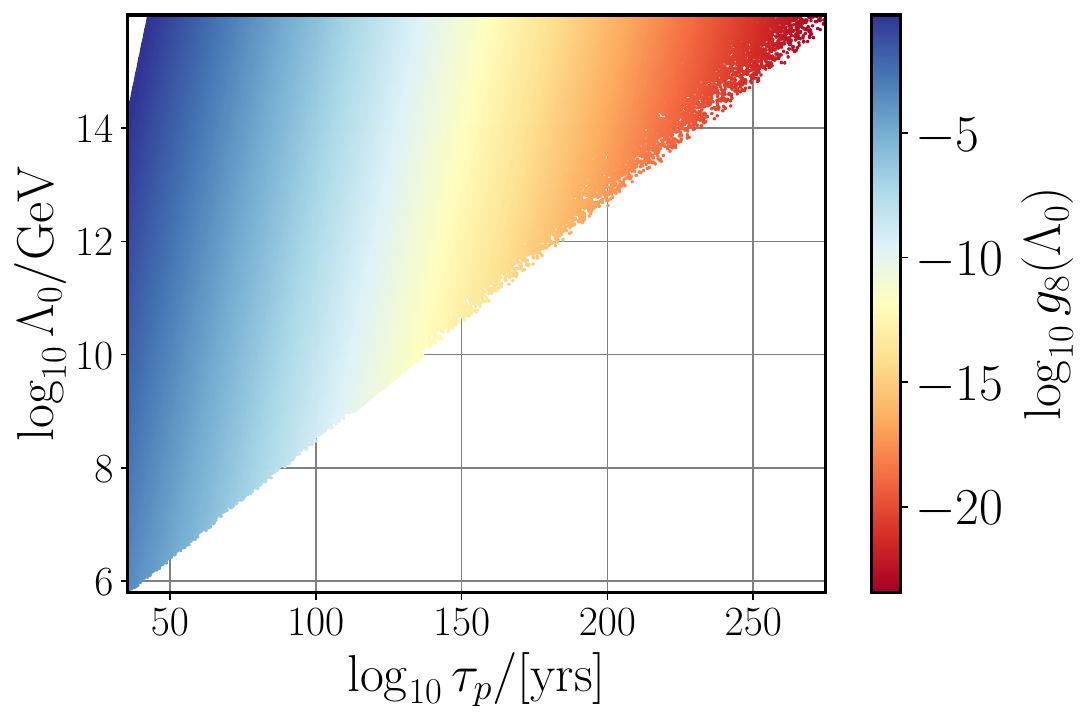}
		\caption{Unification scale ($\Lambda_0)$ and proton lifetime ($\tau_p$). Color indicates the values of the $E_8$ 
		unified gauge coupling $(g_8)$. See~\cite{Aranda:2021eyn} for further details.}
		\label{fig:protondecay}
\end{figure}

For a global visualization of the latter phenomenon, we show the validity domain of the model in Fig.~\ref{fig:protondecay} where, using Eq.~\eqref{eq:prodec} the unification scale $\Lambda_0 \equiv M_\mathrm{KK}$ is represented in terms of the proton lifetime $\tau_p$ and, in the colour scale, the unified $E_8$ gauge coupling $g_8(\Lambda_0)$. The filled region also obeys the constraint in Eq.~\eqref{eq:conspd}, and one takes $l=0.01$ as of the typical size of the loop suppression factor. Notice that Fig.~\ref{fig:gunification} represents the particular case of the bottom-left end of the shaded region where the unification scale finds its smallest allowed value. This result shows that the low-scale asymptotic unification picture can be tested in the proton decay experiments in the future.

\section{Summary}
\label{Sect:summary}

This Letter presents a model featuring the ultimate $E_8$ unification at low-energy scales that provides a viable scenario for phenomenological studies. Its starting point is simple: one gauge group with one representation containing all fields, both matter and forces. This implies the use of $E_8$, ${\cal N}=1$ SUSY, and a ten-dimensional spacetime with six dimensions being compactified on an $\mathbb{T}^6/(\mathbb{Z}_6\times \mathbb{Z}_2)$ orbifold. The orbifold rotational and translational (Wilson line) boundary conditions break the $E_8$ gauge symmetry in anomaly-free way down to the SM gauge group leading to a viable effective low-energy model, to be supplemented with a suitable parametrisation of SUSY breaking effects at low scales. Remarkably, satisfying the proton lifetime constraint and asymptotic gauge coupling unification, leads to a minimal possible unification scale of about $10^{6}\rm{GeV}$. Given these results, and considering that such a scale for unification might be experimentally reachable in the foreseeable future, and further more elaborate effort of the community will be required to explore rich opportunities offered by this proposal.

{\bf Acknowledgements}
The authors thank M\'onica F. Ram\'irez for her support in generating the orbifold diagrams. This work has been supported in part by the MICINN (Spain) projects FIS2011-23000, FPA2011-27853-01, FIS2014-52837-P, FPA2014-53375-C2-1-P, Consolider-Ingenio MULTIDARK CSD2009-00064, SNI-CONACYT (M\'exico), by the Swedish Research Council grant, contract number 2016-05996, by the European Research Council (ERC) under the European Union's Horizon 2020 research and innovation programme (grant agreement No 668679), and by the FCT (Portugal) projects CERN/FIS-PAR/0021/2021, CERN/FIS-PAR/0019/2021, CERN/FIS-PAR/0024/2021, CERN/FIS-PAR/0025/2021 and PTDC/FIS-AST/3041/2020. This work is also supported by the Center for Research and Development in Mathematics and Applications (CIDMA) through the Portuguese Foundation for Science and Technology (FCT), references UIDB/04106/2020 and, UIDP/04106/2020, and by national funds (OE), through FCT, I.P., in the scope of the framework contract foreseen in the numbers 4, 5 and 6 of the article 23, of the Decree-Law 57/2016, of August 29, changed by Law 57/2017, of July 19.


\begin{thebibliography}{100}

\bibitem{Georgi:1974sy}
H.~Georgi and S.~L.~Glashow,
Phys. Rev. Lett. \textbf{32} (1974), 438-441
doi:10.1103/PhysRevLett.32.438

\bibitem{Fritzsch:1974nn}
H.~Fritzsch and P.~Minkowski,
Annals Phys. \textbf{93} (1975), 193-266
doi:10.1016/0003-4916(75)90211-0

\bibitem{Lerche:1986cx}
W.~Lerche, D.~Lust and A.~N.~Schellekens,
Nucl. Phys. B \textbf{287} (1987), 477
doi:10.1016/0550-3213(87)90115-5

\bibitem{Ibanez:1987sn}
L.~E.~Ibanez, J.~E.~Kim, H.~P.~Nilles and F.~Quevedo,
Phys. Lett. B \textbf{191} (1987), 282-286
doi:10.1016/0370-2693(87)90255-3

\bibitem{Parr:2020oar}
E.~Parr, P.~K.~S.~Vaudrevange and M.~Wimmer,
Fortsch. Phys. \textbf{68} (2020) no.5, 2000032
doi:10.1002/prop.202000032
[arXiv:2003.01732 [hep-th]].

\bibitem{Aranda:2021eyn}
A.~Aranda, F.~J.~de Anda, A.~P.~Morais and R.~Pasechnik,
Nucl. Phys. B \textbf{993} (2023), 116266
doi:10.1016/j.nuclphysb.2023.116266
[arXiv:2107.05495 [hep-ph]].

\bibitem{King:2005my}
S.~F.~King, S.~Moretti and R.~Nevzorov,
Phys. Lett. B \textbf{634} (2006), 278-284
doi:10.1016/j.physletb.2005.12.070
[arXiv:hep-ph/0511256 [hep-ph]].

\bibitem{Buchmuller:1985rc}
W.~Buchmuller and O.~Napoly,
Phys. Lett. B \textbf{163} (1985), 161
doi:10.1016/0370-2693(85)90212-6

\bibitem{Koca:1982zi}
M.~Koca,
Lect. Notes Phys. \textbf{180} (2005), 356-359
doi:10.1007/3-540-12291-5\_55

\bibitem{Slansky:1981yr}
R.~Slansky,
Phys. Rept. \textbf{79} (1981), 1-128
doi:10.1016/0370-1573(81)90092-2

\bibitem{ArkaniHamed:2001tb}
N.~Arkani-Hamed, T.~Gregoire and J.~G.~Wacker,
JHEP \textbf{03} (2002), 055
doi:10.1088/1126-6708/2002/03/055
[arXiv:hep-th/0101233 [hep-th]].

\bibitem{Ibanez:1986tp}
L.~E.~Ibanez, H.~P.~Nilles and F.~Quevedo,
Phys. Lett. B \textbf{187} (1987), 25-32
doi:10.1016/0370-2693(87)90066-9

\bibitem{Forste:2005rs}
S.~Forste, H.~P.~Nilles and A.~Wingerter,
Phys. Rev. D \textbf{72} (2005), 026001
doi:10.1103/PhysRevD.72.026001
[arXiv:hep-th/0504117 [hep-th]].

\bibitem{Aranda:2020noz}
A.~Aranda, F.~J.~de Anda and S.~F.~King,
Nucl. Phys. B \textbf{960} (2020), 115209
doi:10.1016/j.nuclphysb.2020.115209
[arXiv:2005.03048 [hep-ph]].

\bibitem{Aranda:2020fkj}
A.~Aranda, F.~J.~de Anda, A.~P.~Morais and R.~Pasechnik,
Universe \textbf{9} (2023) no.2, 90
doi:10.3390/universe9020090
[arXiv:2011.13902 [hep-ph]].

\bibitem{Aranda:2020zms}
A.~Aranda and F.~J.~de Anda,
Int. J. Mod. Phys. A \textbf{36} (2021) no.15, 2150112
doi:10.1142/S0217751X21501128
[arXiv:2007.13248 [hep-ph]].

\bibitem{Marcus:1983wb}
N.~Marcus, A.~Sagnotti and W.~Siegel,
Nucl. Phys. B \textbf{224} (1983), 159
doi:10.1016/0550-3213(83)90318-8

\bibitem{Dreiner:1997uz}
H.~K.~Dreiner,
Adv. Ser. Direct. High Energy Phys. \textbf{21} (2010), 565-583
doi:10.1142/9789814307505\_0017
[arXiv:hep-ph/9707435 [hep-ph]].

\bibitem{Dienes:1998vg}
K.~R.~Dienes, E.~Dudas and T.~Gherghetta,
Nucl. Phys. B \textbf{537} (1999), 47-108
doi:10.1016/S0550-3213(98)00669-5
[arXiv:hep-ph/9806292 [hep-ph]].

\bibitem{ParticleDataGroup:2022pth}
R.~L.~Workman \textit{et al.} [Particle Data Group],
PTEP \textbf{2022} (2022), 083C01
doi:10.1093/ptep/ptac097

\bibitem{Wells:2004di}
J.~D.~Wells,
Phys. Rev. D \textbf{71} (2005), 015013
doi:10.1103/PhysRevD.71.015013
[arXiv:hep-ph/0411041 [hep-ph]].

\bibitem{Antusch:2014poa}
S.~Antusch, I.~de Medeiros Varzielas, V.~Maurer, C.~Sluka and M.~Spinrath,
JHEP \textbf{09} (2014), 141
doi:10.1007/JHEP09(2014)141
[arXiv:1405.6962 [hep-ph]].

\bibitem{Murayama:1994tc}
H.~Murayama and D.~B.~Kaplan,
Phys. Lett. B \textbf{336} (1994), 221-228
doi:10.1016/0370-2693(94)90242-9
[arXiv:hep-ph/9406423 [hep-ph]].

\bibitem{Zyla:2020zbs}
P.~A.~Zyla \textit{et al.} [Particle Data Group],
PTEP \textbf{2020} (2020) no.8, 083C01
doi:10.1093/ptep/ptaa104

\bibitem{Takhistov:2016eqm}
V.~Takhistov [Super-Kamiokande],
[arXiv:1605.03235 [hep-ex]].

\end{thebibliography}
\end{document}